\documentclass[prd,preprint,superscriptaddress]{iopart}
\usepackage{graphicx}
\usepackage{iopams}
\usepackage{cite}

\newcommand{\be}{\begin{equation}}
\newcommand{\ee}{\end{equation}}
\newcommand{\ben}{\begin{eqnarray}}
\newcommand{\een}{\end{eqnarray}}

\begin{document}

\title[]{Observational constraints on the free parameters of  an interacting Bose-Einstein gas  as a  dark-energy model}

\author{Hiram E. Lucatero-Villase\~{n}or and  Germ\'{a}n Izquierdo\footnote{ E-mail address:
gizquierdos@uaemex.mx}}
 \address{Facultad de Ciencias, Universidad Aut\'onoma del Estado de M\'exico, Toluca 5000, Instituto literario 100, Estado de M\'exico, M\'exico}
\author{Jaime Besprosvany}
\address{Instituto de F\'{\i}sica, Universidad Nacional Aut\'onoma de M\'exico, Apartado Postal 20-364, Ciudad  de M\'exico 01000,  M\'exico}

\begin{abstract}
Dark energy  is modelled by a  Bose-Einstein gas of  particles with  an attractive  interaction. It  is  coupled to cold dark matter, within a flat universe,  for the  late-expansion  description,   producing variations in particle-number densities. The model's parameters, and physical association, are: $\Omega_{G0}$,  $\Omega_{m0}$,  the dark-energy rest-mass energy density and the dark-matter term scaling as a mass term, respectively;  $\Omega_{i0}$,  the self-interaction intensity; $x$,   the   energy exchange rate. Energy conservation relates such parameters.  The Hubble equation   omits  $\Omega_{G0}$,  but also contains $h$, the present-day expansion  rate of the flat Friedman-Lem\^aitre-Robertson-Walker metric,  and $\Omega_{b0}$, the baryon energy density,  used as a prior. This  results in  the four effective  chosen parameters  $\Omega_{b0}$, $h$, $\Omega_{m0}$,  $\Omega_{i0}$, fit with   the Hubble expansion rate $H(z)$, and data from its  value today, near distance,  and supernovas.
We derive   wide $1\sigma$ and  $2\sigma$    likelihood  regions  compatible with definite positive total CDM  and   IBEG mass terms. Additionally, the best-fit value of parameter $x$ relieves the coincidence  problem, and a second potential coincidence problem  related to the choice of $\Omega_{G0}$.

\end{abstract}

\section{Introduction}
The observed accelerated expansion of the Universe suggests that the source with the greatest contribution to the total energy density of the homogeneous Friedman-Lema\^\i tre-Robertson-Walker (FLRW) metric is an unknown form of energy with negative pressure called dark energy (DE)\cite{ade2016planck}. The second source with the greatest contribution is pressureless cold dark matter (CDM), followed by ordinary (baryonic) matter. Both dark sources presumably only interact with ordinary matter and radiation through gravitation, but it is reasonable to assume that there is an interaction between  the dark sources,  as it  avoids or  relieves the coincidence problem \cite{besprosvany2015coincidence},  associated  to DE as a cosmological constant/scalar field. In addition, galaxy-cluster  dynamics data\cite{abdalla2009signature} and the integrated Sachs-Wolfe effect\cite{olivares2008integrated} support the idea of an interacting dark energy (IDE). Several IDE models exist in the literature,  as those motivated by particle physics, thermodynamics,  and phenomenological assumptions\cite{copeland2006dynamics,Bamba2012,Wang2016dmdeinteract}.

The non-relativistic Bose-Einstein gas (IBEG) model,  applied to the late universe,  is an example of an  IDE with a detailed microscopic description. Its attractive interaction component  between the particles  generates a negative pressure. In an application to the early universe\cite{iz2010acc}, it leads to   an accelerated expansion under certain conditions. Including a general energy-exchange mechanism, the IBEG explains the present accelerated expansion of the Universe, and  avoids (or greatly relieves) the coincidence problem. The background dynamics of the IBEG model is similar to that of the $\Lambda$CDM model for certain free-parameter  choices \cite{besprosvany2015coincidence}.

Observational data can constrain the free parameters of the DE models by means of Bayesian methods\cite{verde2010statistical}. Several IDE models have been tested in the literature, e.g.,\cite{duran2010observational,ferreira2016strategies,yang2017constraining}. Some observational data are model independent such as the local Hubble constant measurements\cite{Riess20162,ligo2017gravitational}, the data from the history of the Hubble constant $H(z) \,vs \, z$ obtained by the cosmic chronometer approach\cite{moresco2012improved, stern2010cosmic, moresco20166, moresco2015raising} or the type Ia supernova\cite{riess1998observational,perlmutter1999measurements}. Other data sets are model dependent, as the anisotropy of cosmic microwave background measurements\cite{spergel2007three,ade2016planck}, the baryonic acoustic oscillation (BAO) data\cite{BOSS}, the gas mass fraction\cite{allen2008improved}, and the evolution of the growth function\cite{gong2008growth}. In this paper we find   bounds on the free parameters of the IBEG model by means of observational data that is model independent. Additionally, we compare the observational bounds with other constraints of theoretical origin. This kind of analysis is of vital importance to the IBEG model given the physical nature and detailed information of the free parameters.

The paper is organized as follows. In section \ref{inter}, we review the IBEG model of late accelerated expansion, defining the corresponding Hubble constant $H(z)$. In section \ref{constraints}, we obtain the observational constraints on the model's  free parameters   by means of Bayesian methods. The observational data sets we use are the local Hubble constant measurements, and  the    Hubble-parameter  history data, including type Ia supernova luminosity-distance data. In section \ref{implications}, we compare the observational constraints with  theoretical bounds. Finally, in section \ref{conc}, we summarize the work. From now on, we assume units for which $c=\hbar=k_B=1$. As usual, a zero subindex refers to a variable's present value; likewise, we normalize the present day FLRW metric scale factor by setting $a_{0} = 1$.

\section{Interacting Bose-Einstein gas as dark energy}
\label{inter}

The   IBEG model consists of a  Bose-Einstein gas  of spin-zero non-relativistic particles interacting through an attractive contact local   potential between particle pairs. The quasiparticle
energy is composed of a kinetic and a potential-energy  terms,  and
this feature is maintained for the total energy, where the kinetic contribution is
independent of the particle number  (see \cite{iz2010acc,besprosvany2015coincidence} for a detailed model description).
 If the IBEG is set at a temperature below the critical  one  $T_c$,  particles start becoming  condensate,  with $n_c$ accounting  the condensate number density, and  $n_{\epsilon}$, $n=n_c+n_{\epsilon}$, the non-condensate  and   total number densities, respectively. The gas energy density   then reads \cite{iz2010acc}
\be \rho_g= m n+ \varepsilon^{5/3}(128g)^{-2/3}m^{-1}
n_{\epsilon}^{5/3}+\frac{1}{2}v_0 n^2, \label{densBEG}\ee
where $m$ is  the IBEG particle  mass, providing    the energy density at rest, $\varepsilon= 3 \zeta  (\frac{5}{2} )/(2 \zeta (\frac{3}{2} )) \simeq 0.77$, $g=1$ for spin-zero particles and is related to the kinetic energy term of the gas, and $v_0$ is a negatively-defined  potential term that accounts for the interaction between the particles. The gas pressure can be obtained from its  description in the thermodynamical limit and, using $p=-( {    \frac{\partial E } {\partial V }   }) _{N,S}$,  taking the form

\be p=\frac{2}{3}A n_{\epsilon}^{5/3}+\frac{1}{2}v_0n^2, \label{p_beg}\ee
where $A=\varepsilon^{5/3}(128g)^{-2/3}m^{-1}$.

Although the IBEG gas by itself can be used in a early acceleration cosmological scenario \cite{iz2010acc}, it is necessary to assume that it is coupled to CDM and that the number density is not constant in order to obtain a late acceleration solution in a FLRW frame\cite{besprosvany2015coincidence}. In this case, the model is defined in an expanding flat FLRW metric as
\be
ds^{2}=-dt^{2}+a(t)^{2}\left[ dr^{2}+r^{2}d\Omega_{\theta}
^{2}\right],\nonumber
\ee
where $a(t)$ is the scale factor and $t$, $r$ and $\Omega_{\theta}$ are the time, the radius and solid-angle comoving coordinates of the metric, respectively. The energy-momentum source is composed of  pressureless baryonic matter   with energy density $\rho_b$,  pressureless cold dark matter (CDM)  with energy density $\rho_{dm}$ (Ref. \cite{besprosvany2015coincidence} uses $\rho_{m}$ for  $\rho_{dm}$,)
and the IBEG   with energy density given by eq. (\ref{densBEG}).
The latter two sources are coupled  in such a way that the IBEG non-condensate number density is changed as a Markoff's process following the phenomenological law
\be
n_{\epsilon}=n_{\epsilon i} a^{-3}+ n_{\epsilon 0}a^{3(x-1)},\label{creationrate}\ee
where $n_{\epsilon 0}$ and $x\leq 1$ are the free parameters that model the coupling, and $n_{\epsilon i}=0$ is assumed (as the  influence of the first component diminishes as $a$ increases).

 Note that the IBEG particles created by the coupling are non-condensated and, additionally, no further condensation occurs at that epoch as $T_c$ scales as $T$ (see \cite{besprosvany2015coincidence} for details). Consequently, it is reasonable to set the IBEG condensate particle  number density to null, and, then, $n=n_{\epsilon}$. The energy-density conservation equation for the three sources read
\ben &\dot{\rho}_b + 3H\rho_b = 0, \label{evdensb}\\
 &\dot{\rho}_{dm} + 3H\rho_{dm} = -Q, \label{evdensm}\\
 &\dot{\rho}_g + 3H(\rho_g + p_g) = Q , \label{evdensg}\een
where the dot means derivation with respect to time, $Q$ is the coupling term. This system is completed with the Hubble equation $H^2=(8 \pi G/3)(\rho_b+\rho_{dm}+\rho_g)$.

Using (\ref{densBEG},\ref{p_beg}) and from the derivation of (\ref{creationrate}), the coupling term $Q$ can be obtained
\be Q=3Hx \left(\rho_{G0}\,{a}^{3x-3}+\frac{5}{3} \rho_{c0}a^{5x-5}+2\rho_{i0}a^{6x-6} \right), \label{int2}\ee
where we have defined the parameters
\be \rho_{G0}=m n_{\epsilon 0}, \qquad \rho_{c0}=A n_{\epsilon 0}^{5/3},\qquad\rho_{i0}= v_0 n_{\epsilon 0}^2 /2 . \label{parameters}\ee
(Ref. \cite{besprosvany2015coincidence} uses $c$ for  $n_{\epsilon 0}$).

Consequently, the  three-source  energy densities   evolve as
\ben \rho_b&=&\rho_{b0}a^{-3},\label{densB}\\
\rho_{dm} &=&\rho_{m0}a^{-3}-\rho_{G0}a^{3(x-1)}+\frac{5x\rho_{c0}}{2-5x}a^{5(x-1)}+\frac{2x\rho_{i0}}{1-2x}a^{6(x-1)}, \label{densM}\\
\rho_g &=& \rho_{G0} a^{3(x-1)}+ \rho_{c0} a^{5(x-1)}+\rho_{i0} a^{6(x-1)}, \label{endesBEG}
\een
where $\rho_{b0}$ is the  baryonic matter present-day energy density, and $\rho_{m0}$ is a present-day CDM term that evolves as a mass term (i.e., the    CDM    energy-density component that scales as $a^{-3}$).

The $x=1$ case should be treated apart. Then, the   non-condensate  particle number  evolves proportionally to the volume and scales as $a^3$ in a FLRW scenario. Consequently, the IBEG energy density is constant, as the $\Lambda$ term in the $\Lambda$CDM model. Similarly,  the IBEG  pressure is  constant with $p_g\ne -\rho_g$, so the coupling term is not null but reads
\be Q=-3H \left(\rho_{G0}+\frac{5}{3} \rho_{c0} +2\rho_{i0}
\right), \label{int2x1}\ee
while the CDM in our model evolves as
\be \rho_m =\rho_{m0}a^{-3}-
\rho_{G0}-\frac{5}{3}\rho_{c0}-2\rho_{i0}. \label{densmx1}\ee
In such aspects, we safely conclude   that the   IBEG-model dynamics differs from that of the $\Lambda$CDM model in the $x=1$ case.

We note  that the present-day CDM total energy density is
\be \rho_{dm0}=\rho_{m0}-\rho_{G0}+\frac{5x}{2-5x}\rho_{c0}+\frac{2x}{1-2x}\rho_{i0} .  \label{rhodm}\ee
The free parameters of the IBEG model are then $\rho_{G0}$, $\rho_{c0}$, $\rho_{i0}$, $\rho_{m0}$, $\rho_{b0}$ and $x$.

The Hubble equation for $H$ can be rewritten in terms of the redshift $z=(1/a)-1$ as
\be
\frac{H^2}{H_0^2}=(\Omega_{b0}+\Omega_{m0})(1+z)^{3}+\frac{2\Omega_{c0}}{2-5x}(1+z)^{-5(x-1)}+\frac {\Omega_{i0}}{1-2x}(1+z)^{-6(x-1)},
\label{Hubb2}
\ee
where $H_0$ is the present-day Hubble constant, and we have defined $\Omega_{a0}=(8 \pi G \rho_{a0})/(3H_0^2)$ with $a=b,m,G,c,i$. The  above Hubble constant does not depend on the free parameter $\Omega_{G0}$. A relation between the free parameters can be obtained from Eq. (\ref{Hubb2}) evaluated at $z=0$
\be
\Omega_{c0}=\frac{2-5x}{2}\left(1-\Omega_{b0}-\Omega_{m0}-\frac {\Omega_{i0}}{1-2x}\right),
\label{Omc}
\ee
and, from (\ref{Hubb2}) and (\ref{Omc}), it is possible to rewrite the equation for $H$ as
\be
H(z)=h\left\{ \begin{array}{c}(1+z)^{5(x-1)}+
             (\Omega_{b0}+\Omega_{m0})\left[(1+z)^{3}-(1+z)^{5(x-1)}\right]   \\ \nonumber
                \qquad +\Omega_{i0}\frac{(1+z)^{-6(x-1)}-(1+z)^{5(x-1)}}{1-2x}
                 \end{array}\right\}^{\frac{1}{2}},
\label{Hubb3}
\ee
where $H(z)$ is given in units of $ {\rm km\, s^{-1}\, Mpc^{-1}}$ and $ h=H_0/(100\, {\rm km\, s^{-1}\, Mpc^{-1}\,})$ replaces $\Omega_{c0}$ as a free parameter of the model. Regarding  the  $x=1$ case again,  we point at  similarities of the IBEG  and  $\Lambda$CDM models in $H(z)$; given the above definition,  for $x=1$  the Hubble factor    is formally identical for both models. Using Hubble-factor related data     in the $x=1$ case   would make  the parameters $\Omega_{c0}$ and $\Omega_{i0}$
indistinguishable for the fit, with the combination $-2\Omega_{c0}/3-\Omega_{i0}$ playing an identical role as the $\Omega_{\Lambda}$ term.

The  validity of the model's free parameters  was discussed in \cite{besprosvany2015coincidence}, and  the analysis  used the total CDM energy density $\rho_{dm0}$ and the total IBEG energy density $\rho_g(a=1)$,   fixed with $\Lambda$CDM model  best-fit  values ($0.24\%$ and $0.72\%$ of the total energy density of the Universe, respectively). Additionally, the Hubble expansion rate was set    at $75{ \rm km\, s^{-1}\, Mpc^{-1}}$, based on supernova data.   These values were set in order to compare     the IBEG  and  $\Lambda$CDM models. On the other hand, in this work, no such parameters are fixed beforehand and the model's validity   is discussed as well.

The observational data directly related to the Hubble constant with $z<z_{dec}=1024$ (as the data of the local value of the Hubble constant, the history of Hubble constant data and   the  type IA supernova data) can be used to constrain the free parameters $h$, $\Omega_{i0}$ and $x$ directly. Parameters $\Omega_{b0}$ and $\Omega_{m0}$ have both the same dependence on $z$, so they cannot be constrained separately by Hubble-constant related observations, unless some priors are used. On the other hand, no information on  $\Omega_{G0}$ can be obtained directly from the Bayesian analysis of this kind of data, as it is absent from the Hubble expansion rate expression (\ref{Hubb3}).

\section{Constraints on the IBEG-model free parameters  by the Hubble constant}
\label{constraints}
In order to constrain  the  model's free parameters, we   use   the Bayesian approach and Monte Carlo Markov Chains (MCMC) integrated within the code  SimpleMC\cite{Simple} developed by A. Slosar and J. Vazquez for\cite{vazquez2015cosmological}.

We define
\be
\chi^2_{sud}=\sum_i \left(\frac{y(x_i \vert \theta_{sud})-D_i}{\sigma_i}\right)^2 \,
\ee for a given set of uncorrelated data (sud) consisting  of $N$ points of the type $(x_i,D_i),\, i\in[1,N]$, and corresponding  to a theoretical function of the type $y(x,\theta_{sud})$ ($\theta_{sud}$ being a list of free parameters of a given model,)
where $\sigma_i$ is the error of the data point $i$. The choice of free-parameter   values that minimize $\chi^2_{sud}$   has the  largest probability to be true, given the sud \cite{verde2010statistical}, and we refer to them as the best-fit  parameter values $\theta_{sud}^{bf}$.

We also  derive the  parameter-space confidence regions: the  $1\sigma$ likelihood   is the   parameter-space  region    around the best-fit value for which any choice of $ \theta_{sud}$ has a $68.3\%$ probability for  the sud be measured; and the  $2\sigma$ likelihood  parameter-space  region  has  a $95.4\%$   probability. These regions are used to constrain  the model's free parameters.

In order to evaluate the fit quality of the best-fit parameter values, it is possible to define the degrees of freedom (dof) of a given sud as the number of data points $N$ minus the number of free parameters and, then, to compute $\chi^2_{sud,red}=\chi^2_{sud}(\theta_{sud}^{bf})/(dof)$. The theoretical function with the best-fit parameter values $y(x,\theta_{sud}^{bf})$ poorly fits the given sud when $\chi^2_{sud,red}>>1$, while $\chi^2_{sud,red}$ of order of unity represents a good fit \cite{verde2010statistical}. We also can compare the fit quality of $y(x,\theta_{sud}^{bf})$ with other theories by comparing the corresponding $\chi^2_{sud,red}$.

In this section we consider three sud related to $H(z)$. Given   that  $\Omega_{b0}$ cannot be determined  with this kind of data, and that it is not  directly related to the IBEG model, we will set it  as a Gaussian prior   with mean value $\Omega_{b0}h^2=0.022$, and   standard deviation  $0.002$, values reported in\cite{cooke2014precision} from   primordial-deuterium abundance. For the rest of the free parameters, we  assume a range   used as a prior by the MCMC as well. $x$ is assumed  $0.85\leq x\leq 1$ in\cite{besprosvany2015coincidence}, so we   use the range in the MCMC.  $\Omega_{m0}$ is related to the CDM mass-like term of the energy density, so it is defined positive and in the range $0\leq \Omega_{m0}\leq 50$.   $\Omega_{i0}$ is related to $v_0$ and  is defined in the range $-300\leq\Omega_{i0}\leq0$. We note that both $\Omega_{m0}$ and $\Omega_{i0}$ are not total energy-density terms and they should not be constrained beforehand to be smaller than unity, as the  $\Omega_{a0}$ parameters of the $\Lambda$CDM model. Finally,   $h$ is taken in the range $0.50\leq h\leq 1$, as it is related to the of the Hubble-constant  local value  $H_0$.

We proceed   to obtain the constraints on the free parameters $x$, $\Omega_{m0}$, $\Omega_{i0}$ and $h$.

\subsection{Constraints from local value of the Hubble constant data: Gaussian priors}
\begin{figure}[h!]
\centering
\includegraphics[width=1\textwidth]{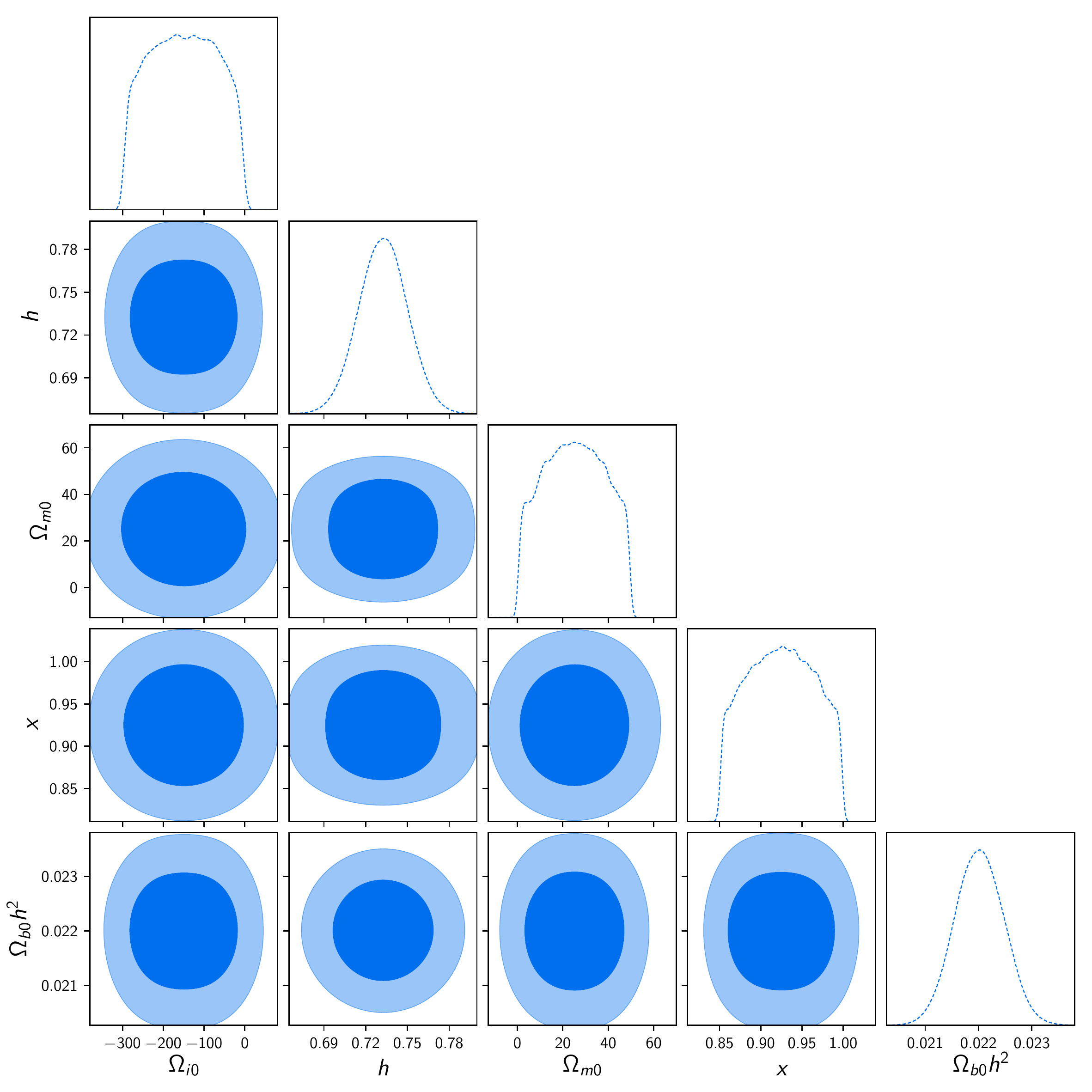}
\caption{Likelihood of the free parameters of the model for the Hubble Space Telescope Gaussian prior.  The darker shaded region corresponds to $1\sigma$, while the lighter shaded region corresponds to $2\sigma$. The plots at the right of the likelihoods represent the free-parameter  probability distribution, with appropriate scale. Similarly for Figures 2-4. }
\label{fig1}
\end{figure}

A  local measurement of the Hubble constant $H_0$ was reported from  the Hubble Space Telescope (HST) data  with a 2.4\% precision in\cite{Riess20162}. The  obtained value is $H_0=73.02\pm1.79 { \rm km\, s^{-1}\, Mpc^{-1}}$. We use it as a Gaussian prior to obtain a first  parameter   constraint.  The result is given in table \ref{Fin} and the $1\sigma$ and $2\sigma$ contours can be appreciated in figure \ref{fig1}.

An independent measurement of the local Hubble constant was recently obtained by the gravitational wave (GW) detectors  LIGO  and Virgo, jointly with the 1M2H, the Dark Energy Camera GW-EM, DES,  DLT40,  Las Cumbres Observatory,     VINROUGE, and   MASTER Collaborations \cite{ligo2017gravitational}. They   measured it  both in the GW spectrum and in the electromagnetic spectrum, from the merger of a binary neutron-star system. The value obtained is $H_0=70^{+12}_{-8}{\rm km\, s^{-1}\, Mpc^{-1}}$, which is compatible with the other measurement considered above to $1\sigma$. We proceed as   previously, setting a Gaussian prior to $h$ with mean value $0.70$ and   standard deviation   $0.08$. The  IBEG   free-parameter  constraints are  presented in table \ref{Fin} under the tag GW. Given that both values considered in this section are compatible to $1\sigma$, the   IBEG  free-parameter   contours observed   are almost identical, with the exception of $h$.

There are other   local Hubble-constant  estimates  including cosmic microwave background (CMB) measurements from Planck\cite{ade2016planck} ($67.8 \pm {\rm 0.9 km\, s^{-1}\, Mpc^{-1}}$), and Barionic Acoustic Oscillation (BAO) measurements from SDSS30, strong lensing measurements from H0LiCOW31, and high-angular-multipole CMB measurements from SPT32. All of them are model dependent via the linear-perturbation evolution and the background dynamics. We   decided not to consider them as we restrict ourselves to measurements related to the background dynamics of the model.

\subsection{Constraints from history of the Hubble-constant data}

We next consider $37$ observational data on the   Hubble-constant history, $H(z)\, vs \, z$, represented in table \ref{HDiagram} with their respective uncertainties and sources in the literature.  The data from\cite{moresco2012improved, stern2010cosmic, moresco20166, moresco2015raising} are obtained by the cosmic-chronometer approach and are model independent. The remaining data are related to BAO observations that can be model dependent. In this case and given the large uncertainties in the measurements, we assume that the deviation of the acoustic peak and distance scale of the IBEG model will not be very different from that of the $\Lambda$CDM model, and thus, the corresponding data can be used as well.

\begin{table} [h!]
\centering
\begin{tabular}{|l|l|l|l|l|l|l|l|}
\hline
 $z$ & $H(z)$ & $\sigma_{H(z)}$ & ref. & $z$  & $H(z)$ & $\sigma_{H(z)}$ & ref. \\
\hline
 0.07 & 69.0 & 19.16 &\cite{zhang2014four} & 0.570 & 100.3 & 3.7 &\cite{cuesta2016clustering}\\

 0.09 & 69.0 & 12 &\cite{simon2005constraints} & 0.593 & 104 & 13 &\cite{moresco2012improved} \\

 0.12 & 68.6 & 26.2 &\cite{zhang2014four} & 0.6 & 87.9 & 6.1 &\cite{blake2012wigglez} \\

 0.17 & 83 & 8 &\cite{simon2005constraints} & 0.680 & 92 & 8 &\cite{moresco2012improved}\\

 0.179 & 75 & 4 &\cite{moresco2012improved} & 0.730 & 97.3 & 7 &\cite{blake2012wigglez} \\

 0.199 & 75 & 5 &\cite{moresco2012improved} & 0.781 & 105 &12 &\cite{moresco2012improved} \\

 0.2 & 72.9 & 29.6 &\cite{zhang2014four} & 0.875 & 125 & 17 &\cite{moresco2012improved} \\

 0.27 & 77 & 14 &\cite{simon2005constraints} & 0.88 & 90 & 40 &\cite{stern2010cosmic} \\

 0.28 & 88.8 & 36.6 &\cite{zhang2014four} & 0.9 & 117 & 23 &\cite{simon2005constraints} \\

 0.32 & 79.2 & 5.6 &\cite{cuesta2016clustering} & 1.037 & 154 & 20 &\cite{moresco2012improved} \\

 0.352 & 83 & 14 &\cite{moresco2012improved} & 1.3 & 168 & 17 &\cite{simon2005constraints} \\

 0.3802 & 83 & 13.5 &\cite{moresco20166} & 1.363 & 160 & 33.6 &\cite{moresco2015raising}\\

 0.4 & 95 & 17 &\cite{simon2005constraints} & 1.43 & 177 & 18 &\cite{simon2005constraints}\\

 0.4004 & 77 & 10.2 &\cite{moresco20166} & 1.53 & 140 & 14 &\cite{simon2005constraints} \\

 0.4247 & 87.1 & 11.2 &\cite{moresco20166} & 1.75 & 202 & 40 &\cite{simon2005constraints} \\

 0.44 & 82.6 & 7.8 &\cite{blake2012wigglez} & 1.965 & 186.5 & 50.4 &\cite{moresco2015raising} \\

 0.44497 & 92.8 & 12.9 &\cite{moresco20166} & 2.34 & 222 & 7 &\cite{delubac2015baryon} \\

 0.4783 & 80.9 & 9 &\cite{moresco20166} & 2.36 & 226 & 8 &\cite{font2014quasar} \\

 0.480 & 97 & 62 &\cite{stern2010cosmic} & & & & \\
 \hline
 \end{tabular}
\caption{Measurements of $H$ vs  $z$. The Hubble constant and the variance are expressed in  ${\rm km\, s^{-1}\, Mpc^{-1}}$ units.}
\label{HDiagram}
\end{table}

\begin{figure}[h!]
\centering
\includegraphics[width=1\textwidth]{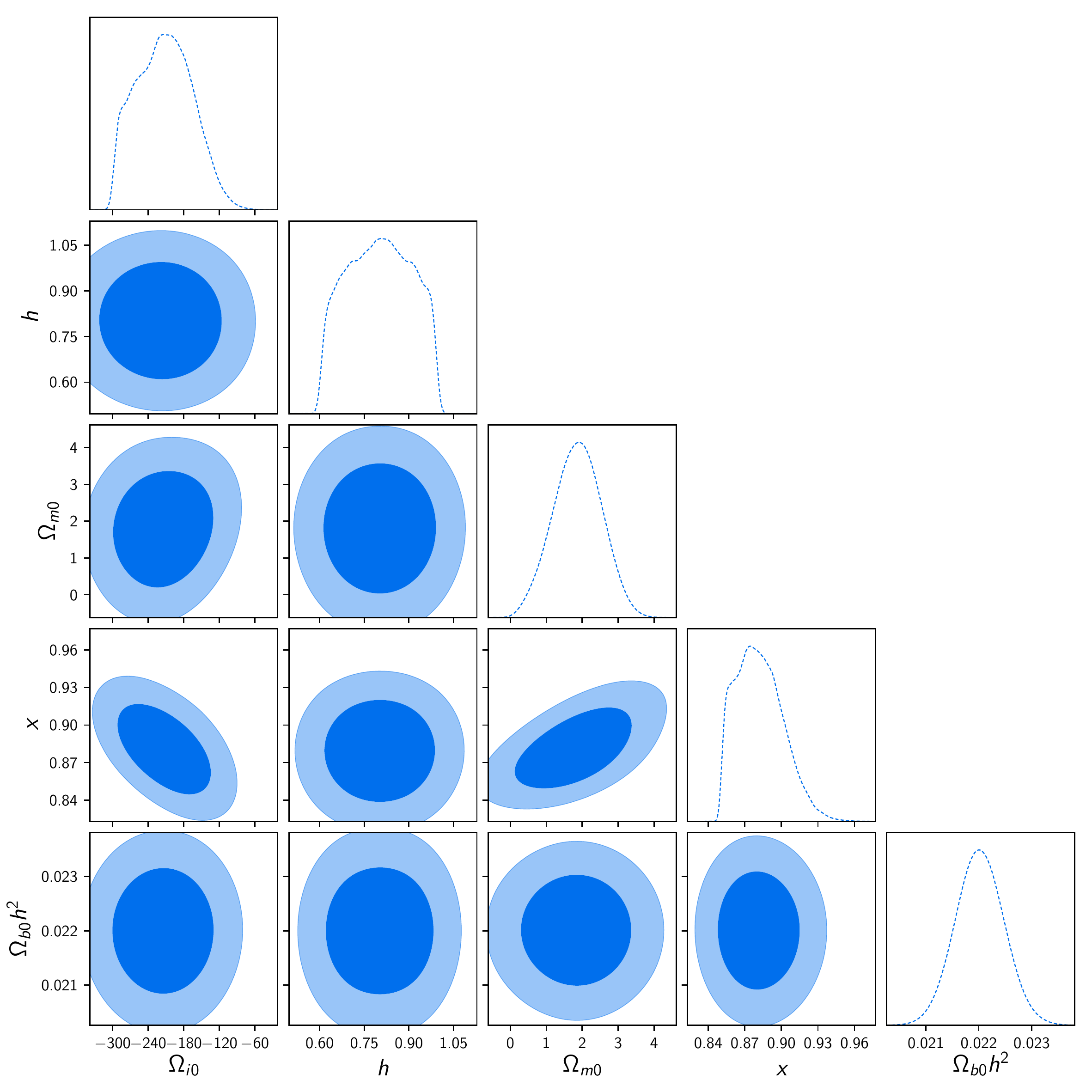}
\caption{Likelihood of the free parameters of the model obtained from the $H(z)$ data.  The darker shaded region correspond to $1\sigma$, while the lighter shaded region correspond to $2\sigma$.}
\label{IntConfHDiagram}
\end{figure}

The constraints on the free parameters from  the $H(z)$ data can be found in table \ref{Fin} and the likelihood is represented in figure \ref{IntConfHDiagram}. The parameter $\Omega_{i0}$ is poorly constricted by this sud, as can be appreciated in \ref{IntConfHDiagram}.

For this sud, the dof$=32$ and, consequently, $\chi^2_{H(z),red}=0.69$ which can be interpreted as the model slightly  'over-fitting' the data. It is possible to use the $H(z)$ data with the $\Lambda$CDM model implemented on the SimpleMC code, and  then, $\chi^2_{sud,red}=1.33$ ($\chi^2_{sud,red}>1$ can be interpreted as the model has not fully captured the data). In this sense, we can only conclude that both the IBEG and $\Lambda$CDM present a similar fit quality of the $H(z)$ data, as both have values  close to $1$.

A comparison between the fit of both models can also be made by means of the  Akaike Information Criterion (AIC) and the Bayesian Information Criterion (BIC) \cite{Liddle04,Liddle07}. The first parameter is defined
\be
AIC=\chi^2(\theta_{sud}^{bf})+2d,
\ee
where $d$ is the  model's number of free parameters. For this sud in the  $\Lambda$CDM case, the parameter is easily computed, $AIC_{\Lambda}=51.84$, while for the IBEG model it is $AIC_{IBEG}=32.08$. The observational data suggest that the "preferred model" corresponds to that  with the smaller AIC parameter. The difference $\Delta AIC=AIC_{\Lambda}-AIC_{IBEG}>10$   is interpreted as strong evidence against the $\Lambda $CDM model. Another criterion is defined by means of the  parameter
\be
BIC=\chi^2(\theta_{sud}^{bf})+d\ln(N),
\ee
where $N=37$ for this sud. Then, $BIC_{\Lambda}=58.28$ and $BIC_{IBEG}=40.13$, and the difference is $\Delta BIC=BIC_{\Lambda}-BIC_{IBEG}>10$, which indicates strong evidence against the $\Lambda$CDM in comparison to the IBEG model.
\subsection{Constraints from Supernova type Ia data.}
The type Ia supernova distance modulus ($\mu$) data is the more constrained model-independent data considered in this work. The distance modulus is directly related to $H(z)$ as
\ben
\mu&=& 5 \log_{10}\left(\frac{d_L}{Mpc}\right)+25 \, ,\nonumber\\
d_{L}&=& \left(1+z\right) \int_{0}^{z} \! \frac{dz'}{H\left(z'\right)}  \, . \nonumber
\een

To get likelihoods of the IBEG parameters from this sud we   use   the joint light curves (JLA) from \cite{betoule2014improved} and a different definition of the $\chi^2_{sud}$, given  that the data is used in 30 correlated bins. In this case,
\be
\chi^2_{JLA}=\sum_{ij} \left( D_i - y(x_i|\theta_{JLA})\right) Q_{ij} \left(D_j-y(x_j|\theta_{JLA}) \right) \, ,
\label{Chicov}
\ee
where $Q_{ij}$ is the $ij$ term of the reported correlation matrix.

\begin{figure}[h!]
\centering
\includegraphics[width=1\textwidth]{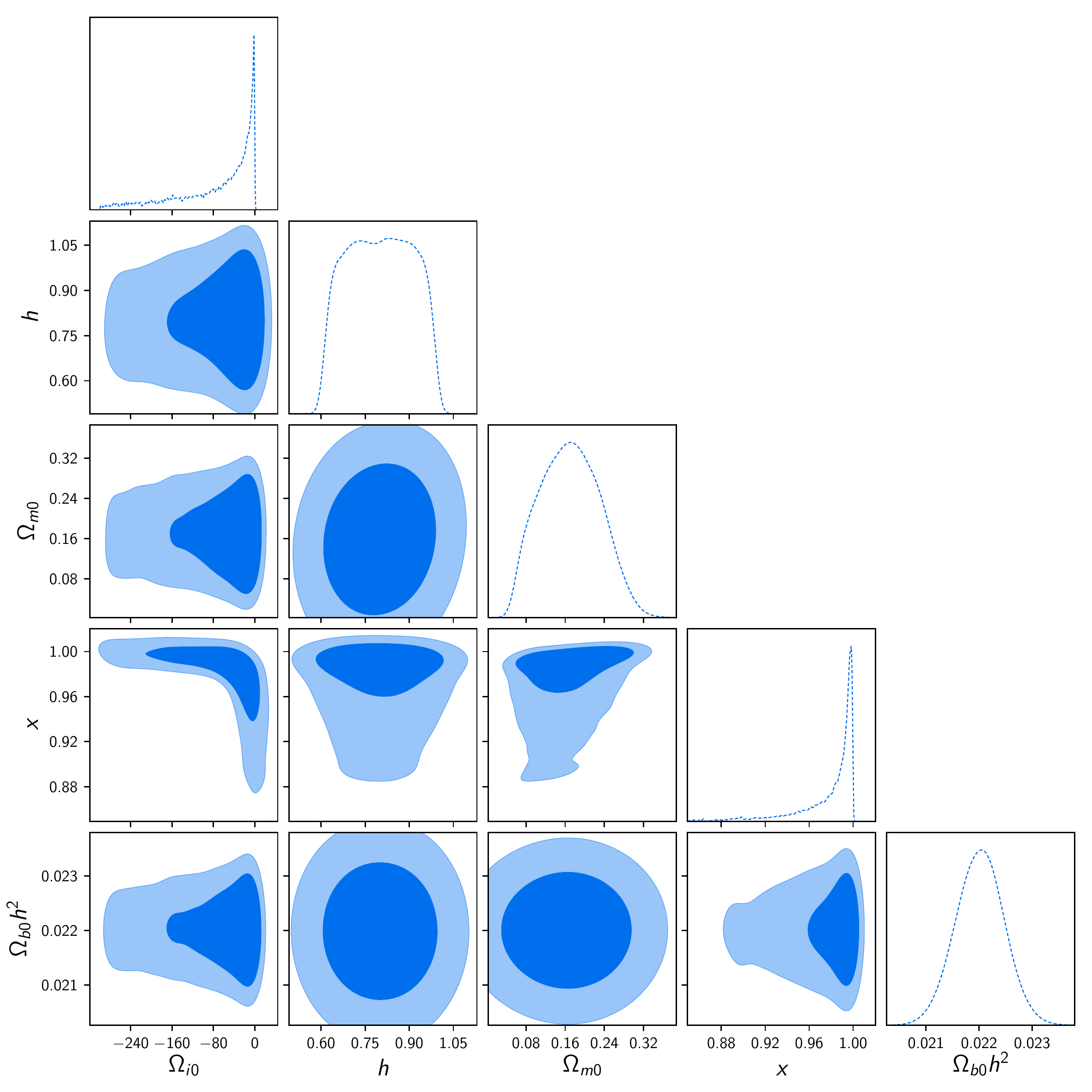}
\caption{Likelihood of the free parameters of the model obtained from the JLA data.  The darker shaded region correspond to $1\sigma$, while the lighter shaded region correspond to $2\sigma$. }
\label{IntConfSN}
\end{figure}

The results of the analysis are found in table \ref{Fin}, and the   likelihood regions in figure \ref{IntConfSN}.  The $1\sigma$ region of  $\Omega_{i0}$ is more restricted from the JLA data than it is in the rest of the data sets. On the other hand, the $2\sigma$ region of the same parameter is still similar to the other data sets.
The dof in this case is $25$ and the corresponding $\chi^2_{JLA,red}=1.30$, which can be interpreted as a good fit of the data. The $\Lambda$CDM model reads $\chi^2_{JLA,red}=1.23$, and we can conclude that both models have an almost identical fit quality.

As in  subsection 3.2, we  compare both models for this sud by means of the AIC and BIC parameters. The AIC parameter of both models is   $AIC_{IBEG}=42.50$ and $AIC_{\Lambda}=39.98$, respectively. The difference between them reads $\Delta AIC=AIC_{IBEG}-AIC_{\Lambda}\simeq 3$ which can be interpreted as   "slight evidence in favor" of the IBEG model. The  parameter   $BIC_{IBEG}=49.5$ and $BIC_{\Lambda}=45.6$ with a difference as $\Delta BIC=BIC_{IBEG}-BIC_{\Lambda}\simeq 4$ is interpreted as "evidence against" the IBEG model. For this sud, both criteria present a tension, which can be explained due to the larger   parameter number of the IBEG model (BIC strongly penalizes the parameter number) \cite{Liddle04,Liddle07}. The tension between AIC and BIC criteria has been reported as well for the $w$CDM model and several interacting DE models \cite{Arevalo17}.
																																																																																																																																																																																																
\subsection{Constraints combining three sets of data: HST$+H(z)+$JLA.}

\begin{figure}[h!]
\centering
\includegraphics[width=1\textwidth]{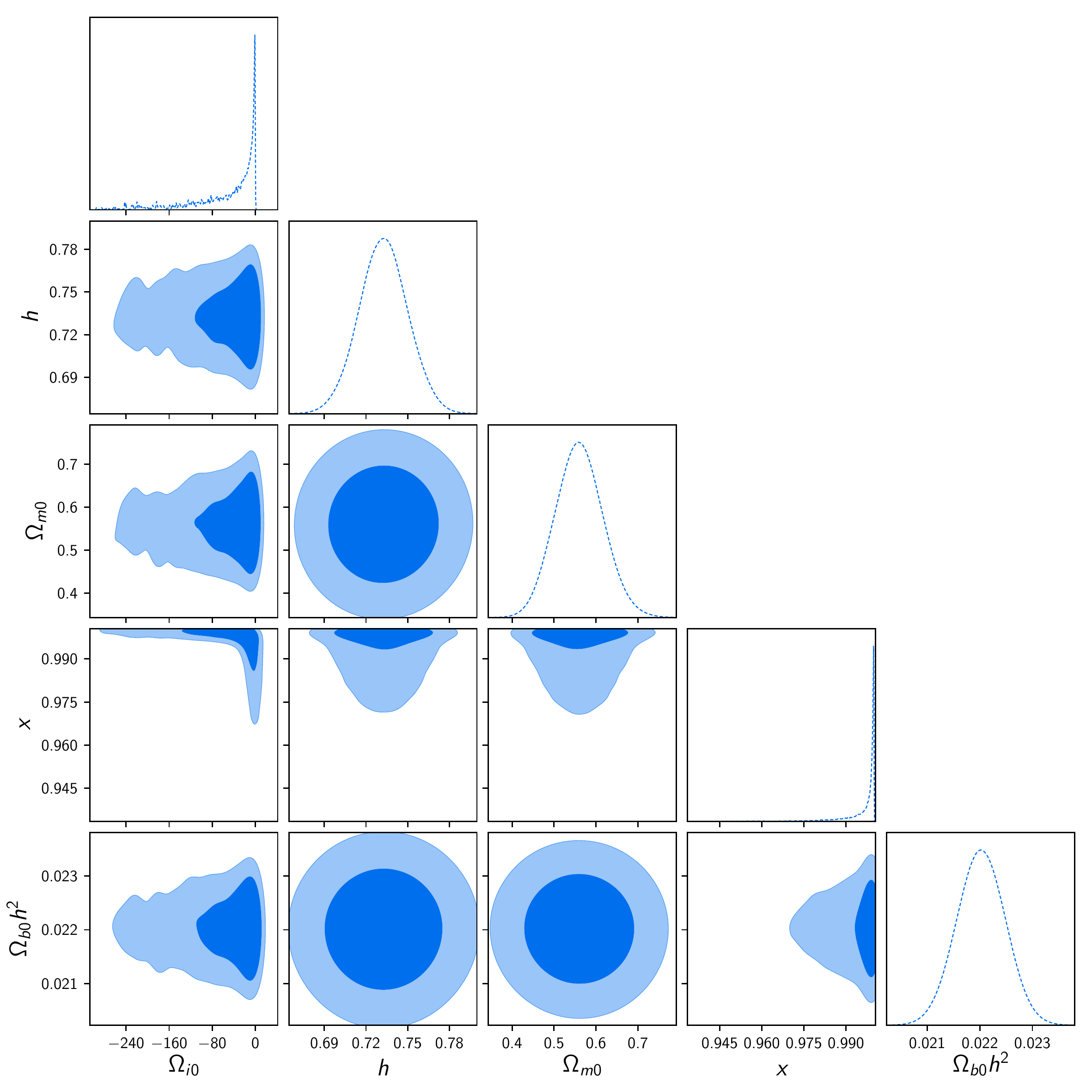}
\caption{Likelihood of the free parameters of the model obtained from the combined HST$+H(z)+$JLA data.  The darker shaded region correspond to $1\sigma$, while the lighter shaded region correspond to $2\sigma$. }
\label{IntConfHD+SN+H0}
\end{figure}

We define
$\chi_T^2=\chi_{H_0}^2+\chi_{HST}^2+\chi_{JLA}^2$ to obtain a total likelihood combining the three sets of model-independent observations and run the MCMC implemented in SimpleMC to explore the   free-parameter space. For the present-day Hubble parameter, we consider the Gaussian prior HST
motivated, on one hand, by the fact that it is the observation of present-day Hubble parameter with the smaller error, and, on the other hand, by the fact that the GW observation is $1\sigma$ compatible with it. The total constraints and likelihoods of the free parameters are shown in table \ref{Fin} and figure \ref{IntConfHD+SN+H0}, respectively.

\begin{table}[]
\centering
\begin{tabular}{l|l|l|l|l|l|l|}
\cline{2-7}
                                    & \multicolumn{2}{c|}{HST} & \multicolumn{2}{c|}{GW} & \multicolumn{2}{c|}{$H(z)$} \\ \hline
\multicolumn{1}{|l|}{Param.}    & Value     & $1\sigma$     & Value     & $1\sigma$    & Value       & $1\sigma$     \\ \hline

\multicolumn{1}{|l|}{$\Omega_{m0}$}    &  14.11  &  14.16    & 47.70  &14.07     &      3.35 & 0.95       \\

\multicolumn{1}{|l|}{$\Omega_{b0}h^2$} &  0.022   & 1$\times10^{-3}$&  0.021 & 1$\times10^{-3}$   &  0.022 & 1$\times10^{-3}$          \\

\multicolumn{1}{|l|}{$h$}           &  0.73  & 0.02      &0.68    & 0.03     &        0.66 & 0.10    \\

\multicolumn{1}{|l|}{$\Omega_{i0}$} &$-$74.72& 47.15    & -247.65 &   57.89   &      $-$233.12 & 65.75    \\

\multicolumn{1}{|l|}{$x$}           & 0.92 & 0.04       & 0.85    & 0.04      &      0.92 & 0.03 \\ \hline
\end{tabular}
\begin{tabular}{l|l|l|l|l|}
\cline{2-5}

                                 & \multicolumn{2}{c|}{JLA} & \multicolumn{2}{c|}{HST+$H(z)$+JLA} \\ \hline
\multicolumn{1}{|l|}{Param.} & Value     & $1\sigma$     & Value           & $1\sigma$         \\ \hline
\multicolumn{1}{|l|}{$\Omega_{m0}$} &    0.10 & 0.07       &                 0.52 & 0.08     \\
\multicolumn{1}{|l|}{$\Omega_{b0}h^2$}   &  0.023&3.0$\times10^{-4}$& 0.022    &  2.0$\times10^{-4}$  \\
\multicolumn{1}{|l|}{$h$}        &    0.62 & 0.05       &                 0.70 & 0.02      \\
\multicolumn{1}{|l|}{$\Omega_{i0}$}&  $-$23.77 & 27.05       &                 $-$3.60 & 12.38               \\
\multicolumn{1}{|l|}{$x$}      &    0.98 & 0.02        &                   0.97 & 0.01 \\ \hline
\end{tabular}
\caption{Constraints on the IBEG free parameters from: the local Hubble constant obtained by HST and GW as  Gaussian priors; history of the Hubble constant $H(z)$; type Ia supernovae (JLA); and combined data (HST+$H(z)$+JLA)  .}
\label{Fin}
\end{table}

\section{Implications of the observational constraints on the IBEG model}
\label{implications}

Next, we compare the observational bounds with theoretical  considerations, e.g., that the  IBEG  energy-density  mass term be positive definite. Given the detailed physical description of the free parameters, the observational constraints could be incompatible with  physical bounds, and the IBEG model could be discarded. Other theoretical demands about the IBEG model are preferable but not compulsory, e.g., that the IBEG model solve or alleviate the coincidence problem.

Most parameters that characterize the IBEG model are constrained and lead to the best-fit values from observations: $\Omega_{i0}=-3.60$;   $\Omega_{c0}=4.79$, where $\Omega_{c0}$ is related to the rest of  the parameters by eq. (\ref{Omc}). However, the model-independent observations related to the Hubble constant, such as $H_0$, $H(z)$ or  JLA , cannot give direct information about  $\rho_{G0}$. Also,  the parameters that define the microscopic nature of the IBEG cannot be  obtained with the above observables (the   IBEG particle mass  $m$, the non-condensate state particle  number    today, $n_{\epsilon 0}$, and  finally, the self-interaction coupling $v_0$.)
The CDM energy density today is also unknown. On the other hand, the energy-exchange term is  constrained by observations to  the best-fit value of $x=0.97\pm0.01$. As stated in\cite{besprosvany2015coincidence},  $x >   0.90$ greatly relieves the coincidence problem in comparison to the $\Lambda$CDM model as $r=\rho_g/\rho_m$ varies at a slow rate compared with the rate of expansion $H_0$. In fact,   $x=1$ solves the coincidence problem. As   $x=1$ is compatible to $2\sigma$ with the observations, we  conclude that its allowed values constrain the IBEG energy-exchange term in a way that they solve  the coincidence problem.

Given that $\Omega_{G0}$ is the IBEG mass term, it is positive definite. The parallelism of  the IBEG and
$\Lambda$CDM models, and observations suggest that $\Omega_{dm0}=(3\rho_{dm0})/(8\pi G H_0^2)$, with $\rho_{dm0}$ defined in (\ref{rhodm}) is also positive, (or, at worst, null.) Then,  using the relation (\ref{Omc}), we   put   the    bound over the free parameters
\ben
\Omega_{dm0}+\Omega_{G0}\nonumber\\
&=&\frac{5x}{2}-\frac{5x}{2}\Omega_{b0}+\frac{2-5x}{2}\Omega_{m0}-\frac{x}{2(1-2x)}\Omega_{i0}\geq 0.\label{thcond}
\een
Although the best-fit values for such parameters, reported in table \ref{Fin} from the HST$+H(z)+$JLA case, is   incompatible with condition (\ref{thcond}),  there is a wide    $ \sigma$ and $2\sigma$  region of compatible likelihoods.

\begin{figure}[h!]
\centering
\includegraphics[width=1\textwidth]{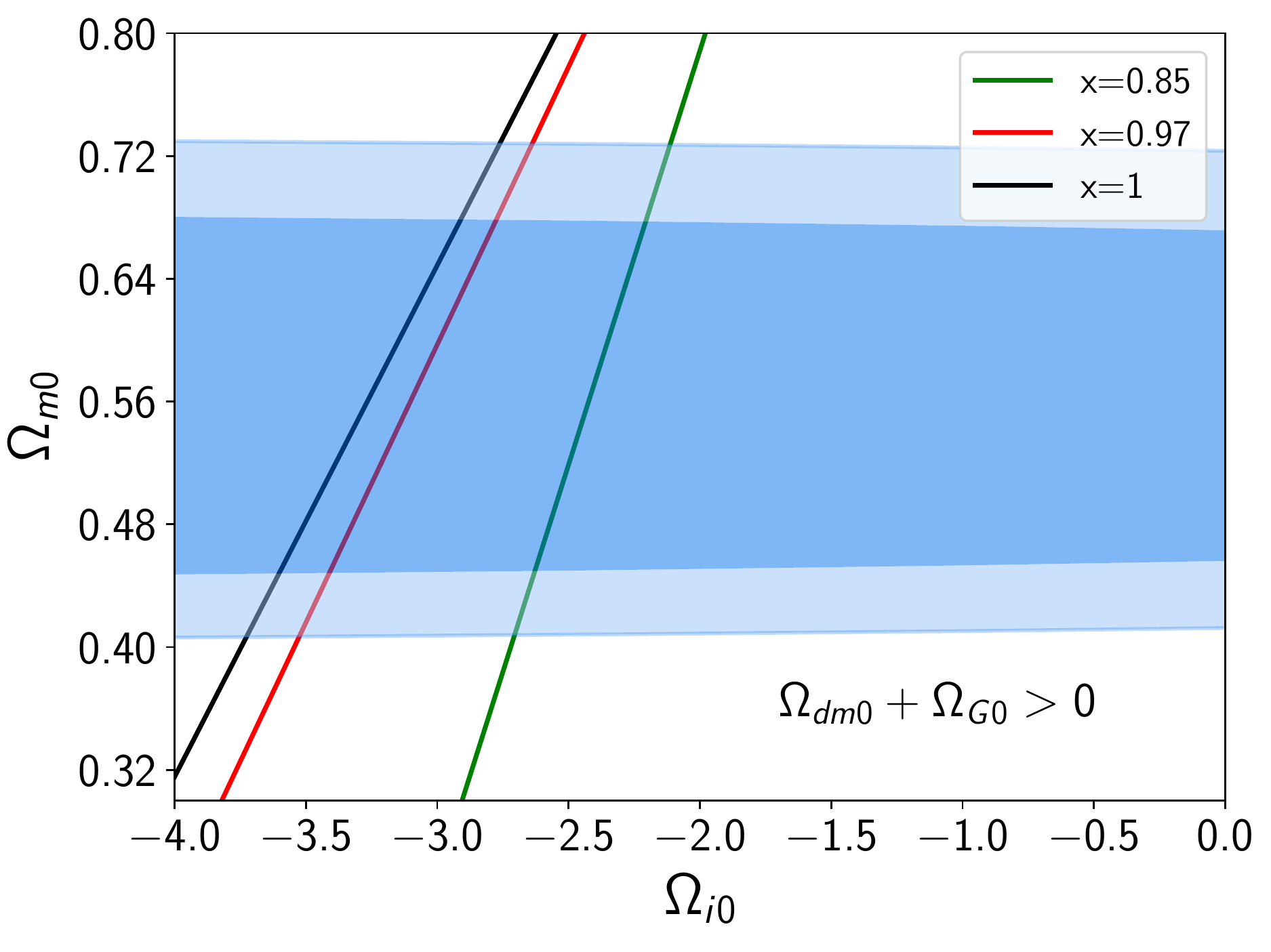}
\caption{Bound $\Omega_{dm0}+\Omega_{G0}=0$ given by (\ref{thcond}) in the $\Omega_{m0} \, vs \, \Omega_{i0}$ space for $\Omega_{b0}h^2=0.022$, $h=0.70$ and three different choices of $x$: $x=0.85$ (green line), $x=0.97$ (red line), and, $x=1$ (black line).   The space on the right-hand side of the line represents   parameter choices   with $\Omega_{dm0}+\Omega_{G0}>0$.  The $1\sigma$ and $2\sigma$ likelihoods of the HST$+H(z)+$JLA case are also shown for comparison.}
\label{fig5}
\end{figure}

Figure \ref{fig5} shows the bound over the parameters given by (\ref{thcond}) for $\Omega_{b0}h^2=0.022$, $h=0.70$, and three different choices of $x$ in the $\Omega_{m0}  \, vs \, \Omega_{i0}$ space, for the $\Omega_{dm0}+\Omega_{G0}=0$, where the space on the right-hand side of the line represents   parameter choices   with $\Omega_{dm0}+\Omega_{G0}>0$. The lines correspond to: $x=0.85$ and $x=1$,   respectively the upper and lower values considered for parameter $x$ in the IBEG model\cite{besprosvany2015coincidence}. Finally, $x=0.97$ is the best-fit value from the HST$+H(z)+$JLA analysis. In the same plot, we show the $1\sigma$ and $2\sigma$ $\Omega_{m0}\, vs \,\Omega_{i0}$ likelihoods from the HST$+H(z)+$JLA analysis. We  conclude that the IBEG model is consistent with condition  \ref{thcond}.

Another, non-compulsory demand for the IBEG model is that  $\rho_g(a_{in})=0$ for some free-parameter choices  and scale factor $a_{in}$ in the past. In\cite{besprosvany2015coincidence}, $a_{in}$ describes the moment when the CDM-IBEG energy interchange starts.

\begin{figure}[h!]
\centering
\includegraphics[width=1\textwidth]{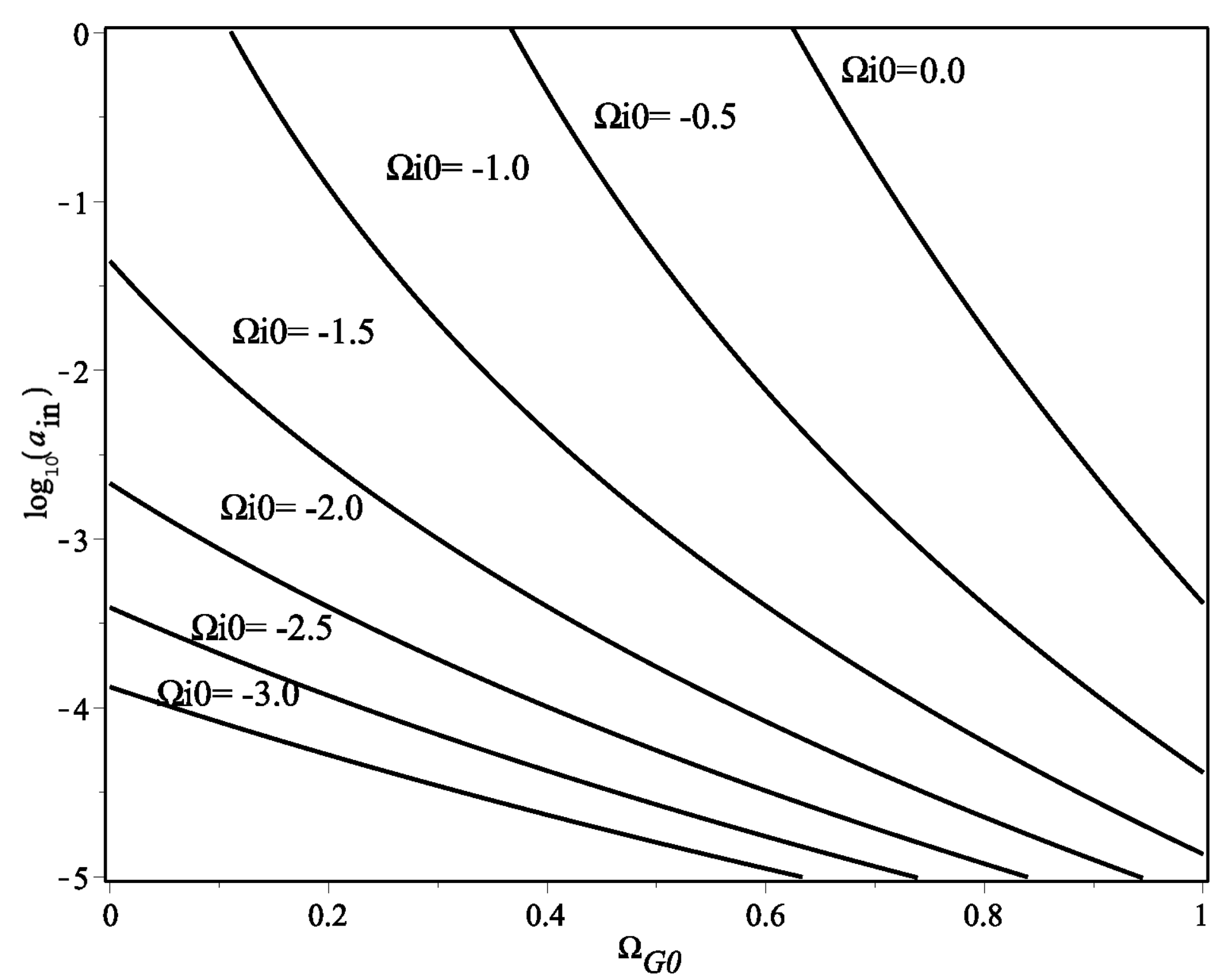}
\caption{The lines represent $\log_{10}(a_{in})\,vs \,\Omega_{G0}$ for $\Omega_{b0}h^2=0.022$, $h=0.70$, $x=0.97$, $\Omega_{m0}=0.52$ and  $\Omega_{i0}$, in the region with $\Omega_{dm0}+\Omega_{G0}>0$.}

\label{ain}
\end{figure}

If $a_{in}$ is near today's value $1$, a second coincidence problem   arises, as  the  acceleration-producing  substance exists only in the near past. Given that $\rho_{G0}$ is not determined by the $H(z)$ observations, it is not possible to limit $a_{in}$ from this work, but assuming the likelihood  contours obtained by the data analysis, there is a wide range of $\Omega_{G0}$ choices that leads to an
$a_{in}$  smaller enough  than $1$. Figure \ref{ain} shows  $\log_{10}(a_{in})\,vs \,\Omega_{G0}$ for $x=0.97$, $h=0.70$, $\Omega_{m0}=0.52$ and different $\Omega_{i0}$, in the region with $\Omega_{dm0}+\Omega_{G0}>0$. For any chosen $\Omega_{i0}$, there is a wide region of $\Omega_{G0}$ for which $a_{in}$ is  smaller enough than $1$, or even smaller than $10^{-4}$. Similar plots can  be obtained for different  $\Omega_{m0}$ inside the likelihood region of figure \ref{fig5}. Despite the lack of information on $\Omega_{G0}$, we  conclude that the coincidence problem associated to $a_{in}$ can be avoided for parameter choices  in the allowed region of figure \ref{fig5}, for $x=0.97$.

\section{Conclusions}
\label{conc}
The IBEG model consists of a flat FLRW metric containing   baryonic,    CDM,  and  IBEG sources, the latter for DE. The CDM,   IBEG components exchange energy at a rate compatible with a Markoff's process. The IBEG particles' attractive self-interaction produces negative presure responsible for an accelerated expansion of the metric\cite{besprosvany2015coincidence}, mimicking the $\Lambda$CDM-model dynamics. The IBEG model parameters have a direct physical  meaning, and any additional bound on them produces information on its constituents. For certain    parameter  choices, the IBEG model solves or alleviates the coincidence problem, and a second related problem, as consistent energy-exchange starts can be found  in the past. This effect is also consistent with the application of the model to the early universe\cite{iz2010acc}, which avoids  the energy-exchange component then.

Observational data on  the background dynamics from the Hubble parameter $H(z)$  constrain the IBEG-model's free parameters. By  setting   limits on their range and comparing with physical requirements, one may maintain or  discard the model. The data used in this work are: the local measurements of the Hubble constant $H_0$\cite{Riess20162,ligo2017gravitational}, the history of
$H(z)$\cite{zhang2014four}-\cite{font2014quasar}, and the modulus distance of type Ia supernova in the joint light curves (JLA) from \cite{betoule2014improved}. A Gaussian prior is used on the baryonic energy density based on universe measured values  \cite{cooke2014precision}.

The data strongly bound the parameter $x$, related to the IBEG energy-exchange rate, finding the best fit  $x=0.97\pm0.01$, which relieves the coincidence problem in the IBEG model. The $1\sigma$ and $2\sigma$ likelihoods for the parameters related to CDM mass energy density ($\Omega_{m0}$) and the IBEG self-interaction ($\Omega_{i0}$) are wide, not too restrictive, and  compatible with the theoretical and observational condition   $\rho_{dm0}+\rho_{G0}\geq 0$. The data bounds are also compatible with  parameter   choices  not affected by the second coincidence problem related to the time when the IBEG-CDM energy exchange starts, defined   in terms of the scale factor $a_{in}$. In this sense, we can conclude that the IBEG model remains consistent with  the observational data used in this work and   Bayesian analysis, which     restrict  the  IBEG-model free parameter  space,  relieving the traditional  and decay-start coincidence problems. The AIC and BIC parameters computed for the history of
$H(z)$ data tend to favor the IBEG model compared to the $\Lambda$CDM model. On the other hand, the JLA data present a tension between both criteria, AIC parameters favoring the IBEG model while BIC one favoring the $\Lambda$CDM model. Such a tension  is present in other DE models (such as the $w$CDM or some interacting DE models), and can be explained in terms of the larger    free-parameter  number of the latter \cite{Arevalo17}. In general, we   conclude that the IBEG model is favoured by the observational data, in the same way as other interacting DE data recently studied in the literature \cite{Wang2016dmdeinteract, Arevalo17, Nunes16, Xia16}.

Finally,  the IBEG particle  mass-related parameter     $\rho_{G0}$ cannot be bounded by the kind of data used in this work. Other observational data, related to the linear perturbation evolution of the model considered, can constrain it. Those model-related observations include the   CMB anisotropy measurements\cite{spergel2007three,ade2016planck}, the BAO data\cite{BOSS}, the gas-mass fraction\cite{allen2008improved}, the evolution of the growth function\cite{gong2008growth}. These data are being analyzed and will be presented in a future work.

\section*{Acknowledgments}
The authors acknowledge   financial support from DGAPA-UNAM,  project IN112916.

\end{document}